\documentclass[aps,prr,reprint,superscriptaddress,footinbib,longbibliography]{revtex4-2}

\usepackage{usebib}
\usepackage{graphicx}
\usepackage{color}
\usepackage{colortbl}
\usepackage{here}
\usepackage{multirow}

\newcommand{\nminv}{nm$^{-1}$}
\newcommand{\gcm}{g cm$^{-3}$}
\newcommand{\vixs}{$v_\mathrm{IXS}$}
\newcommand{\vins}{$v_\mathrm{INS}$}
\newcommand{\vus}{$v_\mathrm{US}$}
\newcommand{\vinf}{$v_{\infty}$}
\newcommand{\vzero}{$v_{0}$}
\newcommand{\sfast}{$S_{f}$}
\newcommand{\szero}{$S(0)$}

\newcommand{\cpp}{$C_{P}$}
\newcommand{\cvv}{$C_{V}$}
\newcommand{\s}{ }
\newcommand{\alphap}{$\alpha_{P}$}
\newcommand{\betat}{$\beta_{T}$}
\newcommand{\deltav}{$\Delta_{v}$}

\begin{document}
\title{Experimental Observation of Mesoscopic Fluctuations to Identify Origin of Thermodynamic Anomalies of Ambient Liquid Water}

\author{Yukio Kajihara}
\email{kajihara@hiroshima-u.ac.jp}
\author{Masanori Inui}
\affiliation{Graduate School of Advanced Science and Engineering, Hiroshima University, Higashi-Hiroshima, Hiroshima 739-8521, Japan}
\author{Kazuhiro Matsuda}
\affiliation{Department of Physics, Kumamoto University, Kumamoto 860-8555, Japan}
\author{Daisuke Ishikawa}
\affiliation{Japan Synchrotron Radiation Research Institute (JASRI) Sayo-cho, Hyogo 679-5198, Japan}
\affiliation{Materials Dynamics Laboratory, RIKEN SPring-8 Center, Sayo, Hyogo 679-5148, Japan}
\author{Satoshi Tsutsui}
\affiliation{Japan Synchrotron Radiation Research Institute (JASRI) Sayo-cho, Hyogo 679-5198, Japan}
\author{Alfred Q. R. Baron}
\affiliation{Japan Synchrotron Radiation Research Institute (JASRI) Sayo-cho, Hyogo 679-5198, Japan}
\affiliation{Materials Dynamics Laboratory, RIKEN SPring-8 Center, Sayo, Hyogo 679-5148, Japan}
\date{\today}

\begin{abstract}
 We report an experimental observation of mesoscopic fluctuations underlying the thermodynamic anomalies of ambient liquid water. 
 The combination of two sound-velocity-measurement methods with largely different frequencies, namely inelastic X-ray scattering (IXS) in the terahertz band and ultrasonic (US) in the megahertz band, allows us to investigate a relaxation phenomenon that has a characteristic frequency between the two aforementioned frequencies.
 We performed IXS measurements to obtain the IXS sound velocity of liquid water from the ambient conditions to the supercritical region of liquid--gas phase transition (LGT) and compared the results with the US sound velocity reported in the literature.
 We found that the ratio of the sound velocities, \sfast, which corresponds to the relaxation strength, obtained using these two methods exhibits a simple but significant change.
 Two distinct rises were observed in the high- and low-temperature regions, implying that two relaxation phenomena exist.
 In the high-temperature region, a peak was observed near the LGT critical ridge line, which was linked to changes in the magnitude of density fluctuation and isochoric and isobaric specific heat capacities.
 This result indicates that the high-temperature relaxation originates from the LGT critical fluctuation, proving that this method is effective for observing such mesoscopic fluctuations.
 Meanwhile, in the low-temperature region, \sfast\s increased from 550 K toward the low-temperature region and reached a high value, attaining the ``fast sound'' state under ambient conditions.
 This result indicates that another mechanism of relaxation exists, which causes the sound velocity anomaly, including the ``fast sound'' phenomenon of liquid water under ambient conditions.
 The change in \sfast\s in the low-temperature region is linked to the change in the isochoric heat capacity, which identified that this relaxation causes the well-known heat capacity anomaly of liquid water.
 This low-temperature relaxation corresponds to the critical fluctuation of the liquid--liquid phase transition (LLT) that is speculated to exist in the supercooled region.
 In this study, both LGT and LLT critical fluctuations were observed, and the relationship between thermodynamics and the critical fluctuations was comprehensively discussed by analyzing the similarities and differences between the two phase transitions.
\end{abstract}


\maketitle

\section{\label{sec:intro} Introduction}
 The thermodynamic behaviors of some ``anomalous liquids'' differ considerably from those of other liquids.
 The most well-known case is of water under ambient conditions.
 The density of liquid water is the greatest at a temperature slightly its melting temperature; the slope of the melting curve against pressure is negative, and the heat capacity exhibits a divergence-like increase in the supercooled region.
 The (ultrasonic (US) or adiabatic) sound velocity increases with the temperature and reaches its maximum value at approximately 350 K.
 Several scenarios have been proposed to explain the thermodynamic anomalies of liquid water; however, a solution has not been determined.
 A liquid--liquid phase transition (LLT) scenario (or a liquid--liquid critical point (LLCP) hypothesis \cite{water_llcp_poole, water_llcp_mishima}) is regarded as the most promising scenario \cite{water_2tale, llt_review_tanaka2020}.
 In this scenario, two structures exist in liquid water with a discontinuous (first-order-like) phase transition between them, which may have its own critical point in the high-pressure and deep supercooled region.
 Ambient water is located in the LLT supercritical region; its thermodynamic features are strongly affected by the LLT ``critical fluctuation,'' which causes anomalies.
 For such an LLT scenario, several theoretical and simulation studies have preceded experimental investigation.
 For instance, Anisimov's group constructed a thermodynamic equation by assuming LLCP and quantitatively reproduced many thermodynamic quantities, such as density, heat capacity, surface tension, and sound velocity (of zero-frequency limit), measured in the liquid region, including the supercooled region \cite{water_thermodynamics_holten2014}.
 However, the estimated critical pressure of LLCP, which is approximately zero, may be extremely low compared with the values estimated by other researchers (approximately 50 \cite{water_llcp_mishima2010, water_llcp_fuentevilla} to 150 MPa \cite{water_confined_llcp, water_llcp_poole}).
Furthermore, some of the estimated properties, such as compressibility, isochoric heat capacity, and sound velocity, exhibit unphysical features near LLCP.
 It is uncertain if a critical point really exists.
 In this sense, a two-order-parameter model proposed by Tanaka may be more appropriate.
 In the model, the existence of two states or two structural orderings is assumed; however, it does not necessarily assume an LLCP or real (first-order-like) LLT between these two orderings.
 By introducing a frustration of the two orders, it is possible to quantitatively explain not only the many thermodynamic anomalies of liquid water \cite{water_2para_tanaka} but also other properties of various liquids, such as glass transition, using simulation techniques \cite{water_review_tanaka2012}.

 To prove the LLT scenario or to solve the problem of the thermodynamic anomalies of water, clear experimental results are necessary.
 However, although more than a quarter of a century has passed since the scenario was proposed \cite{water_llcp_poole} and numerous experiments have been subsequently conducted, conclusive evidence is yet to be obtained \cite{water_2tale}.
 This is mainly because the real first-order-like LLT and LLCP are expected to be located in the deep supercooled ``no-man's land'' \cite{water_llcp_mishima} at high pressure, where real experiments cannot be conducted.
 However, we believe that there is a method to prove the LLT concept without reaching the no-man's land.
 This method involves observing the ``critical fluctuation'' that are believed to be widely spread around the LLCP or LLT and predicting the LLCP from this spread. 
 Analogously, even when the top of Mt. Everest is covered with clouds, if we find a path (a ridge) and follow this path upward, we can move toward the summit \cite{water_saxs_xfel_introduction}.
 In principle, it is not clear what physical quantities are good indexes for LLT critical fluctuation;
 we will discuss and propose the quantities in this paper.
 In the case of the liquid--gas transition (LGT) critical fluctuation, the magnitude of density fluctuation is widely recognized as a good index, which can be measured through small-angle scattering (SAS) experiments using X-rays and neutrons.
 As for the anomalies near ambient conditions of liquid water, SAS measurements have shown that the SAS intensity increases toward low temperatures, including the supercooled region \cite{water_saxs_1st}.
 A hydrogen-bond network \cite{water_saxs_bosio, water_saxs_bosio2}, a clustering \cite{water_saxs_xie}, and recently, the existence of two structures \cite{water_saxs_huang}, have been discussed.
 This interpretation has been controversial for many years \cite{water_saxs_huang_comment}; however, Nilsson's group made a breakthrough using the state-of-the-art X-ray free electron laser (XFEL).
 They succeeded in going into the deep supercooled no-man's land and claimed to have observed the maximum of the integrated SAS intensity, indicating the existence of LLT \cite{water_saxs_xfel}.
 Although caution is still exercised when recognizing this result as the maximum \cite{water_saxs_xfel_comment_caupin}, the rise certainly stagnated.
 To confirm the existence of LLCP or determine the critical pressure, similar SAS measurement in no-man's land at high pressure is desired; however, it is nearly impossible to conduct XFEL experiments under pressurized conditions.
 Alternatively, if critical fluctuation is interpreted as the coexistence of two phases, another experimental verification is possible.
Over a century ago, R\"{o}ntogen claimed that liquid water comprises two different local structures \cite{water_roentgen}; however, it is fundamentally impossible to prove this through normal diffraction measurement \cite{water_struct_difficulty} because the obtained structure factor is spatially averaged. 
 Nevertheless, the notable experimental result on the two structures in liquid water is as follows:
 they are not actual particle structures but electronic structures.
 The oxygen K-edge X-ray emission spectra of liquid water exhibit two distinct peaks \cite{water_xes_fuchs, water_xes_tokushima, water_xes_xas2018} that are believed to correspond to the two structures assumed in the LLT scenario.
 The temperature variations of the intensities of these peaks are consistent with the scenario, which was subsequently extended to the supercooled no-man's land using the XFEL technique \cite{water_xes_xfel}.
 Similar results and discussions have been reported for X-ray absorption spectra \cite{water_xas_science2004_wernet, water_xas_science2004_smith}, although the resolution is not as high as that of X-ray emission spectra.
 These results provide strong evidence of coexistence of two phases; however, each spectrum can be interpreted differently (see \cite{water_spectra_soper} and Section 4.1 in \cite{water_2tale}), and the verification is not conclusive.
 
 Therefore, we focus on another measurement method to detect mesoscopic fluctuations in liquids \cite{kajihara_ixs_te, kajihara_llcp_review_jpn} for replacing the density fluctuation measurement, namely SAS.
 In this method, a phenomenon known as ``fast sound'' \cite{water_fastsound_history, water_fastsound_review_cunsolo} was used.
 The velocity of sound in liquid water under ambient conditions was determined to be approximately 1500 m/s through \textit{macroscopic} methods such as the US technique, and approximately 3000 m/s through \textit{microscopic} methods such as molecular dynamics simulation and inelastic X-ray (IXS) or neutron scattering (INS).
 About the interpretation of this fast-sound phenomenon of liquid water, there have been several conflicting discussions \cite{water_fastsound_history, water_fastsound_review_cunsolo}.
 Also, to this day, there is still debate about microscopic dynamics of liquid water \cite{water_ins_ranieri2016,water_sound_fomin_jml2019,water_sound_fomin_fpe2019,water_sound_fomin2021}, especially the existence of extra excitation mode (transverse-like mode) \cite{ixs_interactmodel2021}, and not all issues have necessarily been resolved.
However, an interesting discussion becomes possible if we look at the \textit{mesoscopic} region between the US and IXS frequencies.
The results of inelastic scattering measurements using ultraviolet (IUVS) and visible light \cite{water_fastsound_iuvs,iuvs_water_licl}, whose characteristic frequencies range from GHz to sub-THz, indicate that the large difference in sound velocities between the different measurement methods is due to the frequency dependence of the sound velocity, and there is a \textit{mesoscopic} relaxation phenomenon with a very large relaxation strength \cite{water_fastsound_review_cunsolo, water_fastsound_iuvs}.
This paper will not go into a discussion of microscopic details of the fast-sound phenomenon, but will discuss the temperature-pressure dependence of this relaxation phenomena in the mesoscopic scales.
 What we focus here is that this fast-sound phenomenon has been observed in systems other than liquid water: 
 In the metal--nonmetal transition (MNMT) region of fluid Hg, inelastic X-ray scattering (IXS) sound velocity is three times faster than US sound velocity \cite{hg_fastsound}, i.e. fast-sound state; 
 Liquid Te, which exhibits thermodynamic anomalies similar to water and is expected to have LLT in the supercooled region, also exhibits 1.7 times fast-sound state near melting temperature \cite{kajihara_ixs_te}. 
 These facts indicate that the mechanism of the fast-sound phenomenon is not the one specific to a material such as ``making (forming) and breaking of hydrogen bonds" \cite{water_fastsound_ixs, water_fastsound_history, water_fastsound_review_cunsolo}, but the one universal and independent of material \cite{water_fastsound_iuvs}. 
 In the study of Hg, the relationship between fast-sound and fluctuation inherent in MNMT was highlighted \cite{hg_fastsound}. 
 In the study of Te, we proposed that the fast-sound phenomenon is a good indicator of critical fluctuation derived from LLT for both Te and water \cite{kajihara_ixs_te}, and later proposed that it can be used for experimental verification of the LLT or LLCP concept \cite{kajihara_llcp_review_jpn}.
 To advance this proposal, we performed IXS measurements on liquid water and discussed the relationship between fast-sound and thermodynamics.
 In the present study, we extend the measurement region not only to the vicinity of the melting point but also to the LGT supercritical region under high-temperature and high-pressure conditions to discuss the LLT critical fluctuation, which is an unknown fluctuation, based on the results of the LGT critical fluctuation, whose properties are well known.
 
 The remainder of this paper is organized as follows.
 Section~\ref{sec:exp} briefly describes the conditions of the IXS experiments. 
 Section~\ref{sec:result} presents the experimental results.
 The focus is on the newly introduced phenomenological parameter ``strength of fast sound" \sfast, which is calculated from the ratio of two sound velocities: the high-frequency sound velocity obtained in this experiment and the low-frequency sound velocity that can be obtained from the literature.
 The remarkable result is that this parameter exhibits a characteristic temperature--pressure variation, particularly in conjunction with that of the isochoric specific heat capacity \cvv.
 In section~\ref{sec:discuss}, we will discuss the experimental result.
 In the first half of the section, the physical meaning of the \sfast\s parameter will be explained, focusing on the relaxation phenomena detected by sound waves.
 In the latter part of the section, the linkage between \sfast\s parameter and \cvv\s will be discussed, focusing on the critical fluctuations of LGT and LLT.
 To make the discussion comprehensive, we will discuss various other data apart from our present experimental data, but to avoid confusion, we will provide details of others' experimental data and literature values in the Appendix.
 Finally, Section~\ref{sec:conclusion} concludes the paper by summarizing our findings and exploring directions for future research.

\section{\label{sec:exp}Experimental}
The experiments were conducted at the high-resolution IXS beamline (BL35XU) of SPring-8 in Japan \cite{sp8_bl35xu_2000}.
 The X-ray conditions were the standard conditions for liquids.
 The energy of the incident X-ray was 21.75 keV according to a Si(11 11 11) back-scattering monochromator, and the scattered X-rays were analyzed using 12 Si analyzers.
 The range of the scattering energy was $-20<E<+20$ [meV] with a resolution of $\Delta E=1.4-1.8$, and the range of momentum transfer was $1.5<Q<11$ [\nminv] with $\Delta Q=0.5$.
 The liquid water sample was introduced into a sample container with a thickness of 4 mm.
The container was made of HASTELLOY X to ensure it endured the strong corrosion effect of supercritical water, and it was equipped with diamond optical windows with thicknesses of 1 mm. 
It is the same as was used for x-ray Raman measurements of supercritical water \cite{samplecell_water_ixs}.
 The container was connected to the sample reservoir and can achieve temperature and pressure up to 800 K and 60 MPa, respectively.
 It was installed into a vacuum chamber equipped with optical windows made of Si with thicknesses of 0.5 mm.
 The measurement was also separately performed for an empty cell without the sample, and IXS spectra of the liquid sample was derived by removing the scattering intensity from the cell by appropriately considering the X-ray transmission of the sample.
  The obtained IXS spectra $I(Q,E)$ were analyzed using the damped harmonic oscillator (DHO) model to assign the energy of longitudinal acoustic mode \cite{kajihara_ixs_te}.
 We define the IXS sound velocity \vixs\s from the slope of the $Q$-dependence of this energy (dispersion curve).

\section{\label{sec:result} Results} 
\begin{figure}[!h]
	\includegraphics[width=80mm]{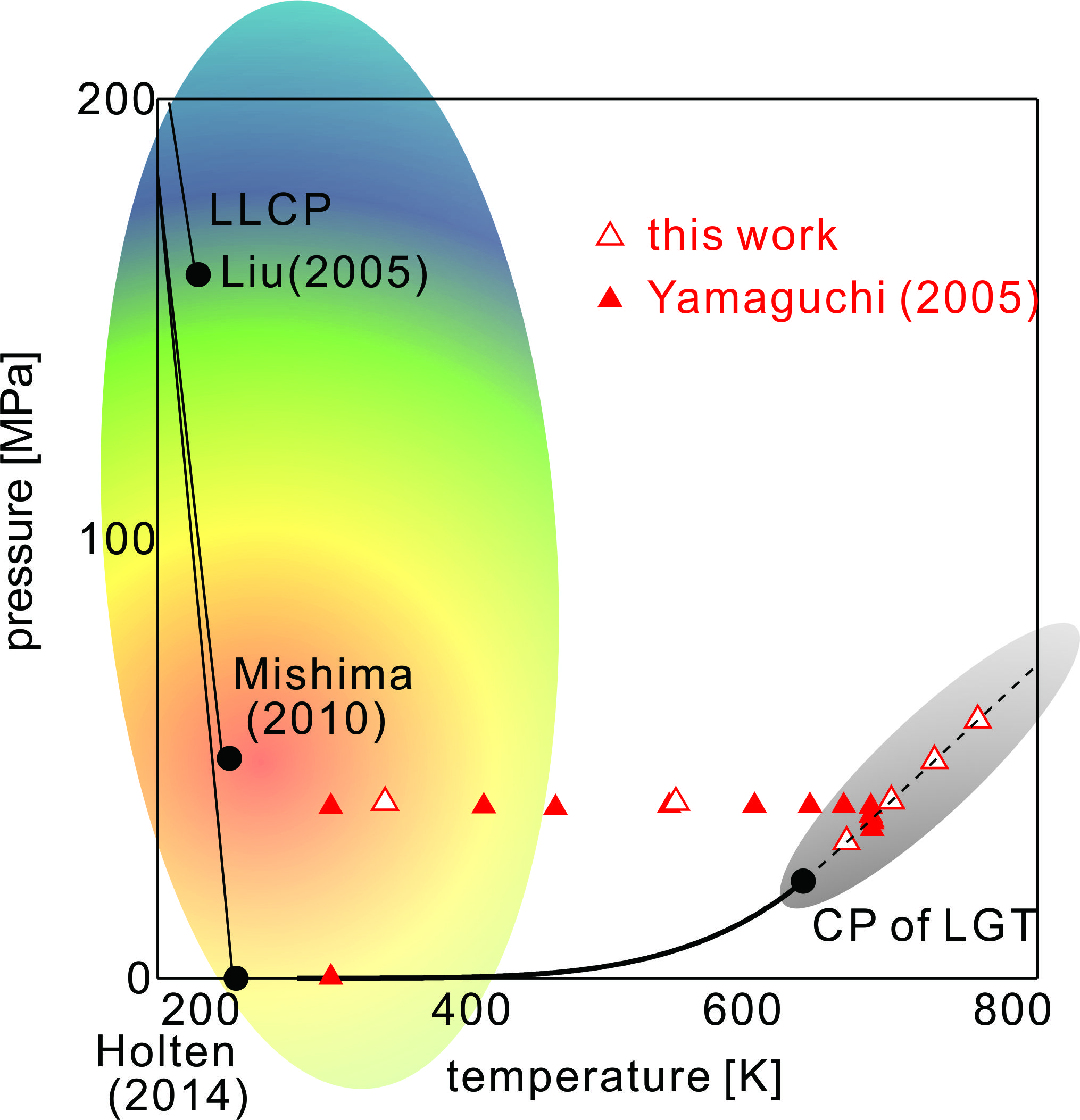}
	\caption{State points of IXS measurements of liquid water in the phase diagram. Our present measurements (open red triangles) and Yamaguchi's measurements \cite{water_ixs_yamaguchi} (closed red triangles) are plotted. CP denotes the critical points of LGT, and LLCPs indicate the critical points of LLT proposed by three representative researchers \cite{water_confined_llcp, water_llcp_mishima2010, water_thermodynamics_holten2014}. The bold and thin solid lines indicate the coexistence line of LGT and hypothetical line of first-order-like LLT, respectively. The dashed line in the high-temperature region is the isochore line of the LGT critical density. The monochrome and colored hatched regions are schematic illustrations of the contour map of the obtained \sfast\s parameter.}
	\label{fig:water_ixs_pt}
\end{figure}

The phase diagram depicted in Fig.~\ref{fig:water_ixs_pt} shows the state points of the measurements obtained from this study and those obtained by Yamaguchi et al. \cite{water_ixs_yamaguchi} with open and closed red triangles, respectively.
 In the phase diagram, CP denotes the critical point of LGT and LLCPs to indicate the critical points of LLT proposed by three representative researchers \cite{water_confined_llcp, water_llcp_mishima2010, water_thermodynamics_holten2014}. 
 The monochrome and colored hatched regions are schematics of the contour map of the obtained strength parameter of fast-sound.
 As can be seen, the state points range from the ambient condition to the LGT supercritical region. 

\begin{figure}[!h]
	\includegraphics[width=80mm]{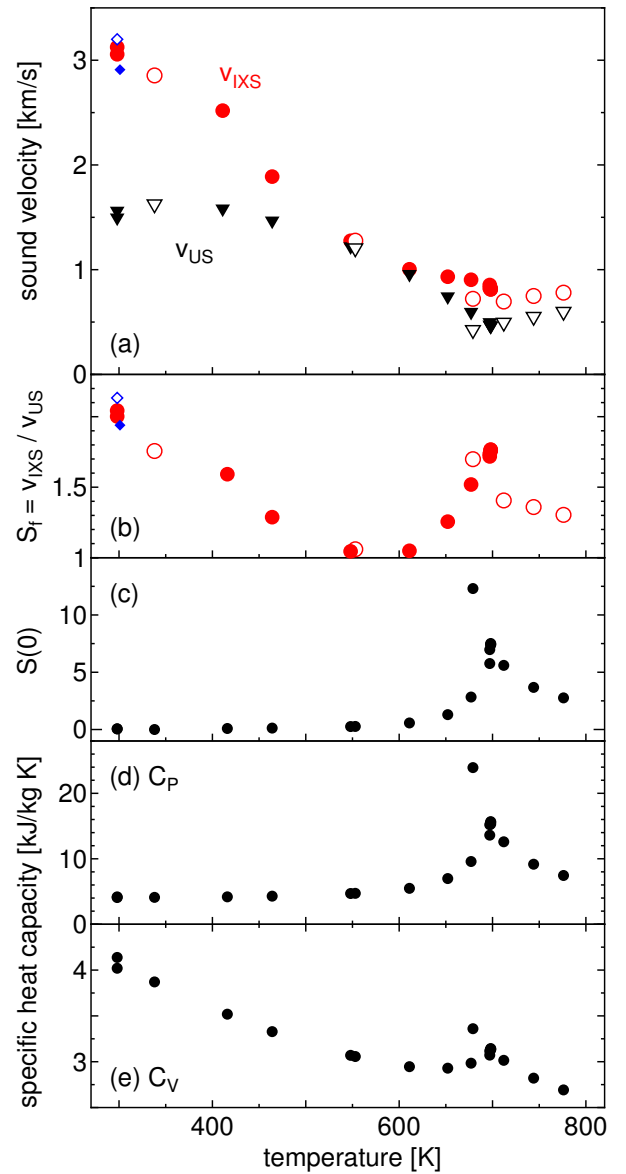}
	\caption{(a) Temperature variation of IXS sound velocities obtained in the present study and those obtained by Yamaguchi et al.\cite{water_ixs_yamaguchi} are represented by open and closed red circles, respectively. IXS sound velocities estimated by other models \cite{water_fastsound_ixs, ixs_interactmodel2021} at ambient conditions are also plotted by open and closed blue diamonds, respectively. The corresponding US sound velocity \vus\s is represented by the triangles. From these two sound velocities, (b) the strength of fast-sound parameter \sfast\s is calculated. Note that the horizontal axis represents the temperature; however, the pressure is not necessarily the same (See Fig.~\ref{fig:water_ixs_pt}). Some thermodynamic parameters of liquid water corresponding to the state points of the measurements are also indicated: (c) zero-wavenumber structure factor \szero\s and specific heat capacities of (d) isobaric \cpp\s and (e) isochoric \cvv. The parameters \vus, \szero, \cpp, and \cvv\s are calculated using the IAPWS-95 formula \cite{water_iapws}.}
	\label{fig:water_params_full}
\end{figure}

 We performed IXS measurements for liquid water under these conditions and deduced the sound velocity from the obtained spectra by adopting DHO analysis \cite{kajihara_ixs_te}.
 The circles in Fig.\ref{fig:water_params_full} (a) represent the temperature variation of the obtained sound velocity \vixs.
 The corresponding US sound velocity \vus\s \cite{water_iapws} is represented by triangles in the same figure.
 The open and closed red circles correspond to our results and Yamaguchi's results, respectively.
 The IXS sound velocities by Sette et al. \cite{water_fastsound_ixs} and Ishikawa and Baron \cite{ixs_interactmodel2021} at ambient conditions are also presented by open and closed blue diamonds, respectively.
 Although they each use a different model in terms of the treatment of another mode in the microscopic level for estimating sound velocity (Sette assumed the existence of a transverse mode, but Ishikawa proposed an improved model that did not introduce this mode and pointed out that the existence of the mode may have been an artifact. See detail in \cite{ixs_interactmodel2021}), the difference in the estimated longitudinal \vixs\s is negligible compared to the difference from \vus, indicating that another mode is irrelevant to the origin of the large difference between \vixs\s and \vus.
 Because the pressure is not necessarily constant, it does not change uniquely with respect to the temperature.  
 Regarding \vixs, although there are small differences between ours and others, the values are nearly consistent and have no significant effect on the following discussion; hence, we will consider them together.
 A very high velocity of approximately 3 km/s was observed near ambient conditions; however, nearly linear deceleration was observed as the temperature increased to approximately 550 K.
 This change is simple and reveals no anomalies.
 At higher temperatures, the change becomes gradual.
 In addition, the red circles at 700 K and higher measured on the critical isochore line indicate that the value remains nearly constant.
 When the density is constant, the sound velocity is also constant, indicating a normal behavior.
 By contrast, the behavior of \vus\s is not normal.
 It rises in the region from 300 to 350 K despite the increase in temperature, and even if the temperature rises to approximately 550 K, the temperature dependence is extremely low.
 Furthermore, \vus\s is nearly equal to \vixs\s around this region.
 At higher temperatures, the temperature dependence becomes relatively steep, and a drop can be observed near the critical isochore line or critical ridge line (700 K at 40 MPa; Figure~\ref{fig:water_ixs_pt}).

 For seemingly contradictory changes in both sound velocities, plotting their ratio or the strength of fast sound \sfast\s$\equiv$\vixs$/$\vus\s in Fig.~\ref{fig:water_params_full}(b) provides a clear indication.
 Specifically, \sfast\s is a relaxation strength detected by sound waves, as will be described in Section~\ref{ssec:relax}.
 In the high-temperature region, \sfast\s exhibits a large peak around the critical ridge line at 700 K.
 Further, \sfast\s rises significantly toward the low-temperature region.
 Then, in the region around 550 K, \sfast\s is nearly 1, i.e., no fast-sound phenomenon occurred.
 This result shows that there are two origins of the observed fast-sound or relaxation phenomena: one is centered at approximately 700 K and the other is in the low-temperature (probably supercooled) region.    
 Figure~\ref{fig:water_ixs_pt} presents a schematic representation of these origins, denoted by the monochrome and colored hatched areas in the phase diagram.
 It is clear that the origin in the high-temperature region is due to the LGT critical fluctuation.
 To confirm this, we present a comparison with other thermodynamic quantities calculated by using IAPWS-95 formula \cite{water_iapws}.
 Figure~\ref{fig:water_params_full} shows (c) the magnitude of density fluctuation $S(0)=\rho_{N} k_{B}T\beta_{T}$ (where $\rho_{N}$ is the number density, $k_{B}$ is the Boltzmann constant, and $\beta_{T}$ is the isothermal compressibility), (d) isobaric (constant pressure) specific heat capacity \cpp, and (e) isochoric (constant volume) specific heat capacity \cvv. 
 In the high-temperature region, (c) \szero\s peaks at approximately 700 K owing to the LGT critical density fluctuation.
 The specific heat capacities (d) \cpp\s and (e) \cvv\s also show peaks at approximately 700 K, similar to \szero.
 These temperature dependences are nearly linked to \sfast, and it is concluded that the fast-sound phenomenon in the high-temperature region is due to the LGT critical fluctuation. 
  
 Meanwhile, in the low-temperature region, \sfast\s shows a substantial increase from approximately 550 K toward 300 K.
 This indicates that there exists another origin distinct from LGT for this fast-sound phenomenon in the low-temperature region.
 The phase diagram in Fig.~\ref{fig:water_ixs_pt} reveals that the phenomenon in the low-temperature region is caused by LLT or LLCP which speculated to exist in the supercooled region.
 Conversely, the increase in \sfast\s is a good indicator of LLT or LLCP hidden in no-man's land.
 The area of the phenomenon of LLT seen in the \sfast\s parameter is considerably wider than that of LGT.
 The difference between the fast-sound phenomenon in the low-temperature region and that of LGT is that the \sfast\s parameter does not seem to be linked with the density fluctuation \szero\s (Fig.~\ref{fig:water_params_full}(c)) and isobaric specific heat capacity \cpp\s (Fig.~\ref{fig:water_params_full}(d)).
 By contrast, the temperature dependence of the isochoric specific heat capacity $C_{V}$ (Fig.~\ref{fig:water_params_full}(e)) suggests that it is also similar to \sfast.
 Moreover, \cvv\s shows a peak at approximately 700 K in the high-temperature region, which implies that it reflects the critical fluctuation of LGT; however, the low-temperature range clearly suggests that it increases with cooling even at 550 K or lower.
 In addition, the absolute value is also larger in the low-temperature range than in the high-temperature range, which is similar to the behavior of \sfast.
 This feature is a new finding that has not been reported previously.

 
\section{\label{sec:discuss} Discussion} 
 In this section, we discuss the physical meaning of the \sfast\s parameter, which was defined phenomenologically from the sound velocity measurements in this study, and provide an interpretation of its linkage with \cvv.
 We will focus on the dual nature of sound velocity.
 The first aspect is as a probe to detect structural relaxation.
 In the first part, we will focus in particular on the time and spatial characteristics of the relaxation phenomena detected by sound waves and show that the parameter corresponds to the relaxation strength at the mesoscopic level.
 The other aspect is as a thermodynamic quantity.
 The (adiabatic or low-frequency) sound velocity is also a thermodynamic quantity, with the equation \vus=$(\rho\beta_{S})^{-1/2}$ between density $\rho$ and adiabatic compressibility $\beta_{S}$ that are thermodynamic quantities.
 In the latter part, we propose to define the \sfast\s parameter as a good index of the critical fluctuations common to both LGT and LLT phase transitions.
 We will use the wide range of temperature-pressure dependence of some thermodynamic quantities as material for discussion, but these are from the literature, not our results, and details will be provided in the Appendix to avoid confusion.
 Integrating these discussions, we will establish that the cause of the thermodynamic anomaly of liquid water is the LLT critical fluctuation and that its entity is relaxation phenomena at the mesoscopic level.

\subsection{\label{ssec:relax} Relaxation phenomena detected by sound wave and mesoscopic fluctuation}
\begin{figure}[!h]
	\includegraphics[width=80mm]{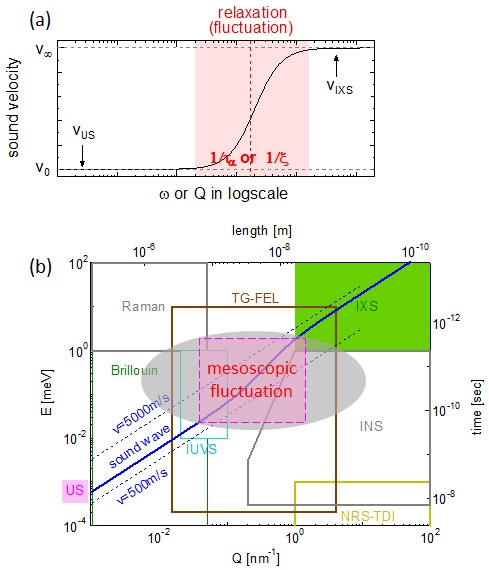}
	\caption{(a)Schematic illustration of the dispersion curve of sound waves. When there exists a certain relaxation or fluctuation with a characteristic time ($\tau_\alpha$) or spatial size ($\xi$) in mesoscopic scales, the measured sound velocity has a high value in the high-frequency region and a low value in the low-frequency region. In general, the frequency of IXS is significantly higher and that of US is significantly lower than the characteristic frequency of the relaxation. We define the ratio of these two sound velocities as \sfast. (b)Energy and momentum transfer (time and length) region of various experimental methods. The dispersion curve is indicated by the solid blue line in the area between the two dashed lines. Mesoscopic fluctuations, which are critical in liquids, exist around the grey ellipse. IXS and US are the largest and the lowest $Q$ and $E$ regions, respectively, in this range, and a large gap exists between them. Light scattering methods (Raman, Brillouin, and IUVS) can fill this gap to a certain extent and free electron laser-based transient grating (TG-FEL) \cite{tgfel, tgfel_nature2015} is expected to fill a considerable part. INS and nuclear resonant scattering using time-domain interferometry (NRS-TDI \cite{tdi_baron1997, saitomakina_2009, saitomakina_prl2012}) are also important for measuring the dynamics of liquids.}
	\label{fig:relax_dynamics}
\end{figure}

 In this subsection, we review the relaxation phenomenon detected by sound waves to understand the physical meaning of \sfast\s introduced in this study.
 Figure~\ref{fig:relax_dynamics}(a) schematically shows the dispersion curve of sound waves.
 Such a diagram was used to interpret the frequency dependence of sound waves for LGT critical fluctuation \cite{critical_phenomena_sound_xe_chynoweth1952}.
 The vertical axis represents the sound velocity while the horizontal axis represents the frequency $\omega$ for the relaxation phenomenon in general. 
 However, in the case of sound waves, the frequency and wavenumber (momentum transfer $Q$ in terms of measurement) are nearly proportional; hence, $Q$ can be used \cite{kajihara_ixs_te,water_fastsound_review_cunsolo}.
 When there is a certain relaxation phenomenon in the system, the velocity will be measured as a higher value \vinf\s at higher $\omega$ (or higher $Q$) and lower value \vzero\s at lower $\omega$ (or lower $Q$), with the inverse of its characteristic time $\tau_{\alpha}$ (length $\xi$) as a boundary.
 LGT critical fluctuation is recognized as relaxation phenomenon observed in such sound waves \cite{critical_phenomena_sound_xe_chynoweth1952}, and we believe that this recognition can be extended to LLT critical fluctuation.
 Alternatively, it is possible to interpret that the sound velocity at low frequency is reduced from that at high frequency owing to the existence of the relaxation phenomenon.  
 We define the dimensionless relaxation strength as follows: 
\begin{equation}
	\Delta_{v} \equiv \frac{v_{\infty}-v_{0}}{v_{0}}.
	\label{eq:strength_relax}
\end{equation}
 Meanwhile, Fig.~\ref{fig:relax_dynamics}(b) shows the energy transfer ($E=\hbar\omega$, where $\hbar$ denotes the reduced Planck constant) and $Q$ region covered by some dynamics measurement methods.
 IXS is located in the highest $E$ and $Q$ region, and it is generally used to observe the local motion of atoms and molecules, whose characteristic frequency is of the order of terahertz.
 By contrast, the frequency of the US wave is in the megahertz region and is located in the lowest $E$ and $Q$ region in this figure.
 The solid blue line in the figure represents the representative dispersion relation of the sound velocity.
 It ranges from approximately 500 to 5000 m/s for most liquids, and if the region between these dotted lines can be measured, the relaxation curve shown in Fig.~\ref{fig:relax_dynamics}(a) will be completed.
 In general, the mesoscopic fluctuation or relaxation region, which is important in disordered systems such as liquids, is located in the region shown by the red ellipse \cite{tgfel}.
 The existence of such a mesoscopic relaxation phenomenon reduces the sound velocity only in the lower $\omega$ (or $Q$) region.
 In general, US has a sufficiently low $\omega$ (or $Q$) and IXS has a sufficiently high $\omega$ (or $Q$) compared to the mesoscopic fluctuation, whose characteristic 
frequency is $1/\tau_{\alpha}$ ($Q$ is $1/\xi$).
 Then, the relaxation strength is approximated as follows: 
\begin{equation}
	\Delta_{v} \simeq \frac{v_\mathrm{IXS}-v_\mathrm{US}}{v_\mathrm{US}} = S_{f}-1.
	\label{eq:strength_relax_szero}
\end{equation}
 Therefore, the strength of fast-sound \sfast\s corresponds to the relaxation strength \deltav. 
 Alternatively, from a different perspective, it can be viewed as a parameter that indicates the degree of reduction in low-frequency sound velocity (due to mesoscopic relaxation phenomena such as critical fluctuations).
 We propose such a relaxation-strength-measurement method (or fast-sound-measurement method based on the historical background) using two sound velocity measurements of different frequencies, namely US and IXS, together, as a new experimental method to characterize liquids in the mesoscopic length and time scales.
 The advantages of this method include the following three points:
 (1) It is simple to take the ratio of \vus\s and \vixs, which does not require complex analysis.
 (2) It can grasp the total strength of relaxations, whose characteristic frequencies range from the megahertz band to the terahertz band.
 The relaxation time of liquid may vary on the order of magnitude with temperature, which cannot be captured by a single experimental technique and requires some techniques.
 Actually, in the fast-sound problem of liquid water, the relaxation time increased rapidly with cooling and soon reached outside the IXS range; thus, the conclusion was reached only when IUVS and Brillouin light scattering were used simultaneously \cite{water_fastsound_iuvs, iuvs_water_licl}.
 The TG-FEL spectroscopy method \cite{tgfel, tgfel_nature2015} covers a wide range of $E$ and $Q$ regions as shown in Fig.~\ref{fig:relax_dynamics}(b), and is a highly effective and ambitious technique; 
 however, it has not been completely developed thus far.
 In any case, it is usually not realistic to measure the entire relaxation curve.
 Even in these situations, the \sfast\s parameter obtained by the relaxation-strength-measurement method is effective.
 Moreover, the shape of the relaxation curve and the number of relaxation phenomena in the frequency range between the megahertz and terahertz bands are irrelevant.
 This is because the relaxation curve in the sound wave is connected by a single curve from the microscopic region to the macroscopic region.
 We believe that this method is highly effective for detecting relaxation phenomena or fluctuations in mesoscopic scales. 
 (3) The last advantage is the scientific value of focusing on relaxation strength.
 Previous dynamics studies in the mesoscopic region of liquids have focused on relaxation times. 
 However, as will be discussed in the next subsection, it is the relaxation strength of sound waves that is directly related to thermodynamics.
 This is expected to develop a discussion that has not received significant attention thus far.

\subsection{\label{ssec:fluct} Relationship between relaxation, thermodynamics, and critical fluctuations}
 In this subsection, we will discuss the relationship between relaxation of the sound waves, critical fluctuations, and thermodynamics.
 First, we review the relationship between sound property and LGT critical fluctuation.
 As shown in Fig.~\ref{fig:water_params_full}(c-e), \szero, \cpp\s and \cvv\s all exhibit maxima near the LGT critical ridge line, implying that these three parameters are good indicators of LGT critical fluctuation.
 In the theory of critical phenomena \cite{book_stanley}, these parameters are said to show the divergent-like increases toward the LGT critical point along liquid--vapor coexistence curve or critical ridge line as \szero$=k_{B}T\beta_{T}\sim|\epsilon|^{\gamma}$, \cpp$\sim|\epsilon|^{\gamma}$, \cvv$\sim |\epsilon|^{\alpha}$ ($\epsilon\equiv(T-T_{c})/T_{c}, \gamma=1.0-1.3, \alpha=0-0.2$), all of which are recognized as good indicators of LGT critical fluctuation from a theoretical perspective.
 It has also been highlighted for sound velocity that it slows down toward the critical point, as $v_\mathrm{US}\simeq v_{S}=(\rho\beta_{S})^{-1/2}\sim|\epsilon|^{-\alpha/2}$ with the same critical index $\alpha$ as \cvv, implying that the degree of reduction of (adiabatic or low-frequency) sound velocity is also a good indicator of LGT critical fluctuation.
However, it has been experimentally found that the degree of this reduction of sound velocity near the LGT critical point becomes smaller as the measurement frequency is increased such as for Xe, CO$_{2}$, SF$_{6}$, $^{4}$He, and $^{3}$He \cite{critical_phenomena_sound_review1970, critical_phenomena_sound_he4_1974, critical_phenomena_sound_he3_1978}.
 This implies that measuring at high frequencies can (partly) freeze LGT critical fluctuation.
 Based on this fact, the relationship between the frequency dependence of complex sound velocity (velocity and absorption) and LGT critical fluctuation or specific heat has been discussed \cite{sound_kawasaki_pra1970,critical_phenomena_sound_co2_pecceu1973}.
 Although not necessarily all of them have been resolved quantitatively, there is already a research direction that considers LGT critical fluctuation as a relaxation phenomenon of sound waves \cite{critical_phenomena_sound_xe_chynoweth1952} and discusses its relationship with specific heats \cite{critical_phenomena_sound_eden1972,critical_phenomena_sound_co2_pecceu1973}.
 The frequency range of the sound measurements in previous studies was significantly limited \cite{critical_phenomena_sound_xe_chynoweth1952} with the upper frequencies in the sub-GHz ranges \cite{critical_phenomena_sound_co2_gammon1967,critical_phenomena_sound_xe_cummins1970,critical_phenomena_sound_sf6_cannell1974}.
 This implies that LGT critical fluctuation with some faster relaxation time was not sufficiently captured by the previous measurements.
 In fact, substantial frequency dependence of sound velocity was observed only in the extreme vicinity of the critical point ($|\epsilon|\lesssim0.01$), where the relaxation time is considered sufficiently slow.
 The frequency band of the IXS experiment conducted this time is THz, which significantly extends this upper limitation.
 It is reasonable to believe that the essence of critical fluctuations is the mesoscopic inhomogeneity or cooperative motion of particles, which has a longer relaxation time than individual particle motions.
 We conclude that the IXS method in the THz band, which detects individual motions, can be recognized as an infinitely high frequency for critical fluctuations.
 The IXS method can completely freeze critical fluctuations and the estimated IXS sound velocity is mostly insensitive to their effects.
 
 In the results of the present experiment (Fig.~\ref{fig:water_params_full}(a,b)), \vixs\s does not slow down at all near the LGT critical ridge line, which is consistent with the aforementioned recognition.
 The fact that substantial sound dispersion (\sfast\s is significantly larger than 1) was observed over a much wider temperature range than before, i.e., $|T-T_{c}|\gtrsim100 K$ or $|\epsilon|\gtrsim0.1$, may be due to the ability to capture smaller LGT critical fluctuation over a wider temperature and pressure range by being able to capture faster relaxation phenomena, as mentioned.
 This fact conversely indicates that the increasing trend of the relaxation strength (\sfast\s parameter) can be used to predict the existence and location of the LGT critical point.
 As a reasonable extension, it is also expected to be used as a new method to detect critical fluctuations associated with other phase transitions as well as LGT \cite{kajihara_llcp_review_jpn}.
 It may be effective to prove the existence of phase transitions and critical points that are experimentally difficult to reach, such as deep supercooled regions and ultrahigh temperature and pressure regions.
 In fact, the present results of \sfast\s (Fig.~\ref{fig:water_params_full}(b)) show an increase toward the low-temperature region, indicating the existence of some other origin of relaxation than LGT in the low-temperature (probably the supercooled) region.
 Considering the aforementioned recognition and its location on the phase diagram (Fig.~\ref{fig:water_ixs_pt}), it seems reasonable to assume that the origin is LLT or LLCP, which has been proposed to exist in the supercooled region, and that the large relaxation strength we observed is an LLT critical fluctuation that extends into the real liquid region above the melting temperature.
 We believe in this conclusion, but we hope that various related experimental data and various discussions will be reported in the future.

 Herein, we propose to use \cvv\s and \sfast\s as good indicators of LLT critical fluctuation.
 In contrast, we conclude that \szero\s and \cpp, which were good indicators of critical fluctuation in LGT, are not good indicators in LLT.
The main reason for this is that we believe there is an essential difference between LGT and LLT as a phase transition;
LGT is a density phase transition in which density is the order parameter \cite{book_stanley}, while LLT is not.
Two of the key impacts of this difference are briefly described below. 
A more detailed discussion using data from others is provided in the Appendix.
(1) The magnitude of density fluctuation differs significantly between LGT and LLT;
 \szero\s in LLT, it is approximately 1/100 or smaller than that in LGT (see Fig.~\ref{fig:s0_cpv}(a)). 
 Thus, for LLT, \szero\s would not be a good indicator of critical fluctuations from a quantitative perspective.
 Additionally, \cpp\s is associated with enthalpy fluctuations (Appendix A in \cite{book_stanley}), but in LGT, the volume effect is by far the largest part of this enthalpy. 
 This overwhelming quantitative difference between LLT and LGT in \szero\s and \cpp\s actually leads to misunderstanding in interpretation under ambient conditions as follows.
 Because \szero\s and \cpp\s appear to increase rapidly in the supercooled region, it is often believed that the density fluctuation anomaly is only in the supercooled region \cite{water_heat_supercool_rasmussen1973} and that the region under ambient conditions is not affected by fluctuations.
 However, considering the aforementioned facts, it can be concluded that near ambient temperature (300 K), there exist two effects of LLT critical density fluctuation, spreading from the relatively close distance from LLCP (from around 220 K \cite{water_llcp_mishima2010} to 184 K \cite{water_llcp_ruishi2020}), and LGT critical density fluctuations with 100 times greater magnitude spreading from the somewhat distant LGT critical point (647 K) (see Fig.~\ref{fig:s0_cpv}(a) inset). 
 It seems reasonable to conclude that the effects of LLT critical fluctuation actually extend to a much higher temperature range above the melting temperature.
 In the end, we concluded that \szero\s and \cpp\s are not good indicators in LLT.
(2) We believe the essence of LLT as an internal-energy phase transition, not a density phase transition.
The importance of the internal energy was discussed in the previous Te paper \cite{kajihara_ixs_te}, showing the actual equation relating \sfast\s parameter (positiveness) and internal energy.
In the case of Te, LLT is accompanied by MNMT, and the valence electronic states are different between the two phases.
This difference in electron energy corresponds to the difference in internal energy.
In the case of water, however, MNMT is not accompanied, of course, but X-ray emission and absorption measurements  \cite{water_xes_tokushima,water_xes_xfel,water_specta2016,water_xas_jpcm2002,water_xas_science2004_wernet,water_xas_science2004_smith} proposed that there are two electronic states of the valence electrons.
 If this is considered a phase transition of internal energy, a common understanding of the LLT of water and Te, two totally different substances, becomes possible.
 \cvv\s is associated with the internal-energy fluctuation (p.130 in \cite{book_statphys_kubo}), which is a reasonable physical quantity as LLT critical fluctuation that accurately represents this essence.
 Additionally, the linkage between \cvv\s and \sfast\s is newly revealed in this study, and we would like to propose \sfast\s as a good indicator of LLT critical fluctuation.
 Because \sfast\s is a parameter that describes the collective particle motions, we may recognize \sfast\s as a measure of dynamical fluctuation in the mesoscopic scales, corresponding to the static fluctuation of \szero\s in the mesoscopic scales \cite{kajihara_llcp_review_jpn}.

Regardless of whether the cause of the relaxation phenomenon is derived from LLT or not, the specific heat capacity \cvv\s and \sfast\s measured in this study are linked, and it is sufficient to conclude that the origin of the heat capacity anomaly under ambient conditions is the relaxation phenomenon of sound waves observed in this study.
 The linkage with \cvv\s is also confirmed by \sfast\s estimates using data reported by others in IXS \cite{water_ixs_krisch2002} and INS \cite{water_ins_ranieri2016} experiments, which are established over a fairly wide temperature and pressure range up to 1 GPa (Fig.\ref{fig:water_fastsound_others}).
 This fact indicates the need to extend the scope of the study beyond the supercooled region to such high-temperature and high-pressure regions in order to elucidate thermodynamic anomalies under ambient conditions. 
Conventional studies on critical fluctuations have focused on the divergent trends of thermodynamic quantities in the extreme vicinity of the critical point. 
 In contrast, it is now clear that the effects of critical fluctuations or relaxation phenomena cover a fairly wide range of temperature and pressure regions and are linked to the changes in specific heats.
 We expect that the wide range of critical fluctuations or relaxation strength, not only in the vicinity of the critical point, will open up a new way to understand the thermodynamics of liquids.
 
 Finally, we would like to present an expression that relates \sfast\s to critical fluctuations.
 The (adiabatic) sound velocity $v_\mathrm{S}$ is related to the adiabatic compressibility as follows: $\beta_\mathrm{S}=1/(\rho v_\mathrm{S}^{2})$.
 Assuming that a critical fluctuation's component $\beta_\mathrm{S}^{fl}$ is added to the normal component $\tilde{\beta_\mathrm{S}}$, it becomes
\begin{equation}
	\beta_\mathrm{S}=\tilde{\beta_\mathrm {S}}+\beta_\mathrm{S}^\mathrm{fl}.
	\label{eq:betas1}
\end{equation}
 The present experimental results are expressed as follows:
\begin{eqnarray}
	\label{eq:betas2}
	\beta_\mathrm{S}=\beta_\mathrm{S}(\omega=0)=\frac{1}{\rho v_\mathrm{S}^2}\simeq\frac{1}{\rho v_\mathrm{US}^2} \\
	\label{eq:betas3}
	\tilde{\beta_\mathrm{S}}=\beta_\mathrm{S}(\omega=\infty)=\frac{1}{\rho v_{\infty}^2}\simeq\frac{1}{\rho v_\mathrm{IXS}^2} \\
	\label{eq:betas4}
	\frac{\beta_\mathrm{S}^\mathrm{fl}}{\beta_\mathrm{S}} 
	 = \frac{\beta_\mathrm{S}-\tilde{\beta_{S}}}{\beta_\mathrm{S}}
	 = 1 - \left( \frac{v_\mathrm{S}}{v_{\infty}} \right)^{2}
	 \simeq 1 - S_{f}^{-2},
\end{eqnarray}
 and the relationship between the critical fluctuation component of adiabatic compressibility and the experimental parameter \sfast\s is obtained.
 Strictly speaking, there is spatial density inhomogeneity in the mesoscopic level, and it may be necessary to distinguish this difference in $\rho$ in Eq.\ref{eq:betas2} and \ref{eq:betas3}, but we neglect it because it is small compared to the difference between \vus\s and \vixs.
 Thus, the relaxation strength \sfast\s is a good index for critical fluctuations.
 We conclude that the proposed relaxation-strength-measurement method is effective as a method for measuring critical fluctuations regardless of LGT or LLT.
 We conclude that the two relaxation phenomena in liquid water that we have measured in the real liquid region (above the melting temperature) are LGT and LLT critical fluctuations.

\section{\label{sec:conclusion} Concluding remarks}
We performed IXS measurements on liquid water over a wide temperature and pressure range to obtain high-frequency sound velocity \vixs\s in the THz band. 
By comparing this value with the low-frequency sound velocity \vus\s in the MHz band in the literature, we extracted the relaxation strength \sfast.
From the experimental result, the relationship between sound wave behavior (relaxation phenomenon) and critical fluctuations is discussed.
The obtained temperature dependence of \vixs\s was monotonically (almost linearly) changing under isobaric conditions and almost constant under isochoric conditions, a ``normal" change that seems to be independent of the existence of LGT and LLT critical points. 
This result is reasonable considering that high-frequency measurement that detects microscopic particle motion is \textit{insensitive} to mesoscopic level of critical fluctuations. 
On the other hand, the relaxation strength \sfast\s in the mesoscopic level showed a distinct increase toward the LGT and LLT critical points, which reveals that \sfast\s is a good parameter as the magnitude of LGT and LLT critical fluctuations.
It has long been known that the temperature-pressure dependence of low-frequency sound velocity \vus\s of liquid water is not monotonic and exhibits ``anomalous" behavior near ambient conditions. 
The present results indicate that the low-frequency sound velocity, that represents macroscopic collective motion of particles, can be interpreted by a combination of two effects:
 the  ``normal" temperature-pressure dependence in the microscopic level and a mountain- or hill-like upward trend toward the LGT and LLT critical points in the mesoscopic level. 
 Table \ref{tbl:relax_method} summarizes the relationship between these sound wave behavior and critical phenomena. 
The relaxation-strength-measurement method, which combines two sound-velocity-measurement methods with largely different frequencies, namely IXS and US, is an effective experimental method for measuring the magnitude of critical fluctuations.
\begin{table*}[htb]
\centering
	\begin{tabular}{c|c|c|c|c}
		\hline
		size & \multirow{2}{*}{exp. method} & \multirow{2}{*}{parameter} & \multirow{2}{*}{feature} & Pressure (P) and Temperature (T) \\
		(frequency)	&                 			     &                      		         &                                 & variation\\
		\hline
		micro        	& IXS  	& \vixs              & insensitive to		& normal           \\
		(THz)        	& (or INS) 	& $\simeq$\vinf  & crit. fluct.      	& (almost linearly with P, T) \\
		\hline
		\multirow{2}{*}{meso} & relax-strength-meas.	 & \sfast  & \multirow{2}{*}{(crit. fluct.)} & hill(mountain)-like \\
		               	             & (IXS + US) 	       & $\simeq\Delta_{v}-1$ &         	                          & (increase toward critical point)\\
		\hline
		macro 	& \multirow{2}{*}{US} & \vus=\vixs$/$\sfast	& sensitive to		& ``anomalous'' \\
		(MHz) 	&								   & $\simeq$\vzero			& crit. fluct.			& (normal + hill(mountain)-like)\\
		\hline
	\end{tabular}
	\caption{Characteristics of relaxation-strength-measurement method and its relation to critical fluctuations.}
	\label{tbl:relax_method}
\end{table*}

The present results also reveal for the first time that the temperature-pressure variation of  the relaxation strength \sfast\s and that of the isochoric specific heat capacity \cvv\s are linked over a wide range. 
This linkage was found to exist not only in the pressure range below 40 MPa by our present experiment, but also in the high pressure range of about 1 GPa by analyzing the results of high-frequency sound velocity measurements by other researchers (see Fig.\ref{fig:water_fastsound_others}). 
In other words, the present results indicate that this relaxation phenomenon detected by sound wave is the origin of the anomaly in specific heat of ambient liquid water. 
We believe it is reasonable to recognize this relaxation phenomenon as LLT critical fluctuation, but to do so, it is necessary to notice the differences from LGT critical fluctuation. 
In the high-temperature LGT critical region, the magnitude of density fluctuation \szero, the isobaric and isochoric specific heat capacities \cpp\s and \cvv, and the relaxation strength \sfast\s all show maxima near the critical ridge line, and these four parameters are all good parameters for LGT critical fluctuation.
In the low-temperature LLT region, however, the changes in \szero\s and \cpp\s are extremely small compared to LGT, and cannot be recognized as good indicators of critical fluctuation.
 We believe that this difference represents an essential part of LGT and LLT as a phase transition.
In LGT, density is an order parameter and is essentially a density phase transition, whereas in LLT, the density (volume) difference between the two phases is extremely small and is not a good variable to describe the phase transition.
As an alternative to \cpp, we would like to propose using \cvv, with volume effects removed, as a good indicator of critical fluctuation.
We would also like to propose that the \sfast\s parameter (or relaxation strength \deltav), which exhibits a temperature--pressure dependence linked to \cvv\s as revealed in this study, be used as a good indicator of critical fluctuation.
This parameter can be said to represent dynamical fluctuation or inhomogeneity in the mesoscopic level, whereas \szero\s represents static fluctuation or inhomogeneity in the mesoscopic level.
 We would like to mention the spread of critical fluctuations in the temperature-pressure region.
 In LGT, steep elongated mountain-like peak is formed along the coexistence curve and critical ridge line.
 On the other hand, in LLT, less directional but large hill of fluctuation are formed over a much wider temperature and pressure region.
Table \ref{tbl:lgt_llt} summarizes the characteristics and parameters of these LGT and LLT critical fluctuations.
\begin{table*}[htb]
	\begin{tabular}{c|c|c|c|c}
		physical quantity & parameter & exp. method & LGT & LLT \\
		\hline
		static fluct. 			& \multirow{2}{*}{\szero}	& \multirow{2}{*}{SAS}  	& \multirow{2}{*}{very large} & \multirow{2}{*}{small} \\
		(density inhomo.)	&											& 										& 	                                   & \\
		\hline
		isobaric heat capacity	& \multirow{2}{*}{\cpp}	& & \multirow{2}{*}{very large}		& exist. \\
		(enthalpy fluct.)				&										&	& 													& (comparable to \cvv) \\
		\hline
		dynamic fluct.			& \multirow{2}{*}{\sfast}	& relax--strength-meas.	& \multirow{2}{*}{large}	& \multirow{2}{*}{large}	\\
		(collective motion)	&											& (sound wave)			& 										& \\
		\hline		
		isochoric heat capacity	&	\multirow{2}{*}{\cvv}	&	& \multirow{2}{*}{exist}	& \multirow{2}{*}{large}	\\
		(internal energy fluct.)	&										&	&										&\\
		\hline
		\hline
		P-, T- spread	&	&	& thin-mountain-like	& wide-hill-like	\\
		\hline
	\end{tabular}
	\caption{Characteristics of LGT and LLT critical fluctuations.}
	\label{tbl:lgt_llt}
\end{table*}
We expect that this unprecedented perspective will lead to new developments in LLT research in the future.
 
 Finally, we would like to mention the scope of application of the relaxation-strength-measurement method.
 We believe that this method can be used as an effective method for characterizing various liquids.
 We have already applied it to water--alcohol mixtures and obtained a direct consequence of thermodynamic anomalies and mesoscopic fluctuations \cite{kajihara_ixs_waterglycerol, kajihara_sound_wateralcohol}.
 As discussed here, it is also a much more sensitive method than the SAS method for detecting mesoscopic inhomogeneity.
 We believe that effective signals can be obtained in high-entropy alloys \cite{hea_original2004, hea_ye2016} and glass-forming liquids, which are believed to have inhomogeneity, but whose signals are not sufficiently obtained by SAS.
 The latter, in particular, is claimed to have "dynamical heterogeneity" \cite{hetero_supercool_ediger2000} and we believe that the property can be obtained directly by this method.

\section*{Author contribution}
The research was organized by Y.K. and the experimental work was executed by all authors, using a vessel procured by the RSC Materials Dynamics Laboratory (A.Q.R.B. and D.I.).  
The paper was written by Y.K. in discussion with K.M. and reflects comments from all authors.

\begin{acknowledgments}
 This work was supported by JSPS KAKENHI Grant Numbers 20244061 and 20K03789. 
 The IXS measurements of liquid water were performed at the Synchrotron radiation facility of Japan (BL35XU of SPring-8) with the approval of the Japan Synchrotron Radiation Research Institute (JASRI) (Proposals No. 2008B1108).
\end{acknowledgments}

\section*{appendix}
In this study, we have developed an unprecedented discussion of the relationship between thermodynamic anomalies in liquid water and relaxation phenomena detected by sound wave by focusing on critical fluctuations at the mesoscopic level.
To avoid confusion, we have limited our discussion in the main text almost exclusively to our experimental data, but it would be very interesting to apply arguments in this direction to data reported by other researchers.
In this appendix, (A) others' experimental results of high-frequency sound velocities \cite{water_ixs_krisch2002,water_ins_ranieri2016}, and (B) the wide range of temperature and pressure dependence of thermodynamic quantities \cite{water_iapws}, will be discussed as materials.
By integrating these results, (C) essential differences between LGT and LLT critical fluctuations will be discussed and provided as a supporting materials for the conclusions in the main text.

\subsection{\label{ssec:velocity_others} High-frequency sound velocity in the high-pressure region by other researchers}
\begin{figure}[!h]
	\includegraphics[width=80mm]{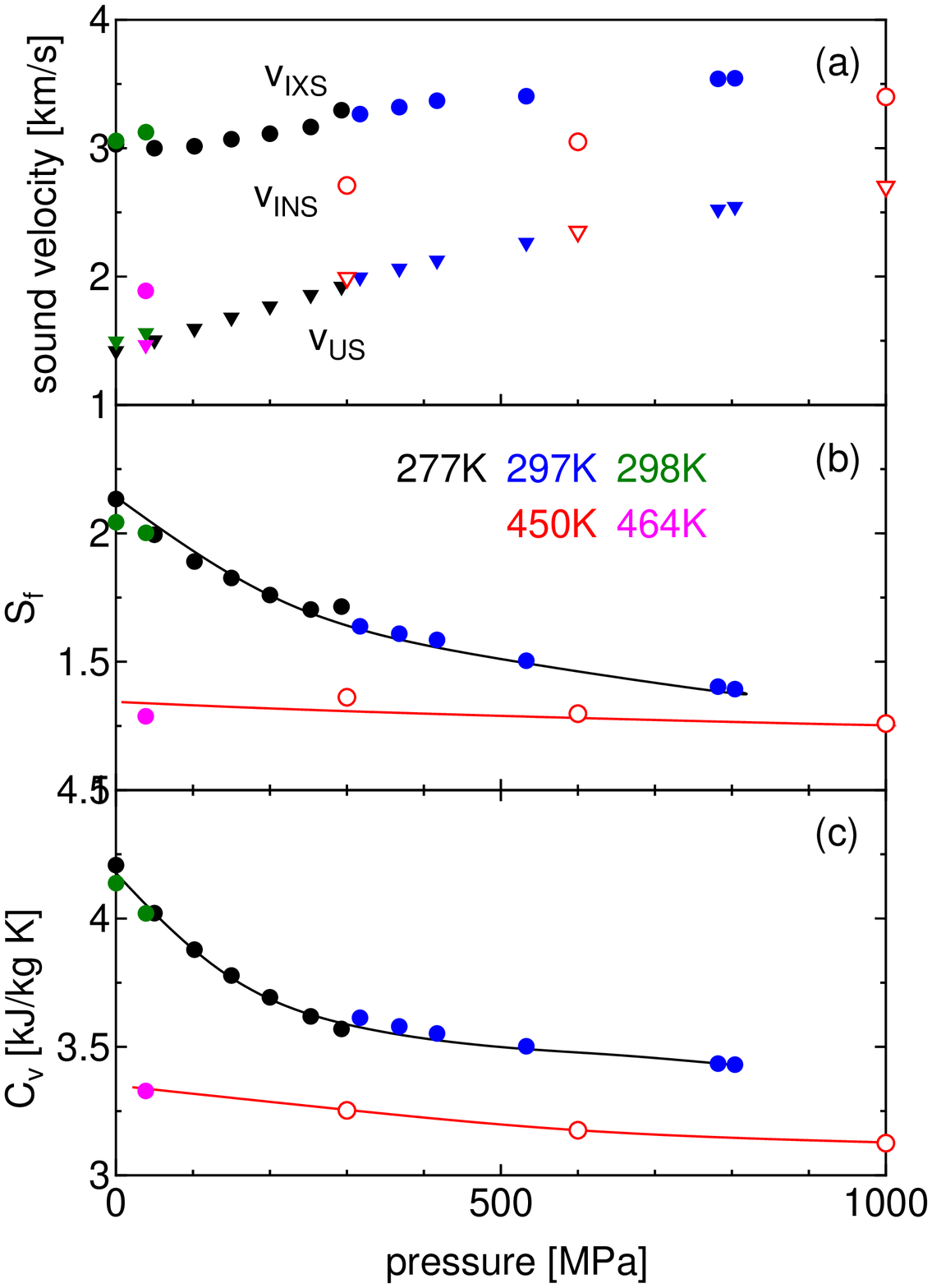}
	\includegraphics[width=85mm]{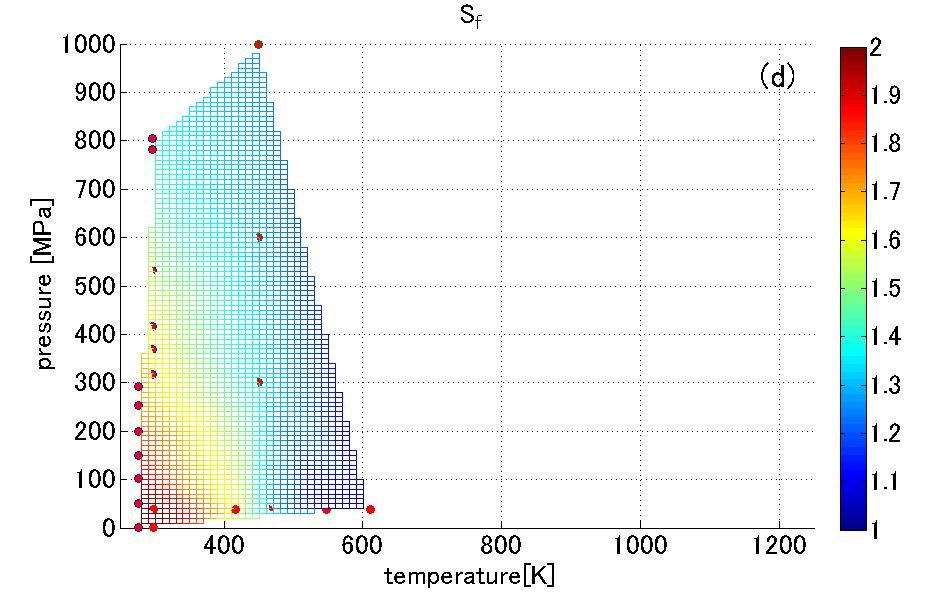}
	\caption{(a) Pressure variation of IXS and INS sound velocities at low-temperature and high-pressure region reported by others indicated by circles (277 K (black symbols) and 297 K (blue) \cite{water_ixs_krisch2002}. 450 K (red) \cite{water_ins_ranieri2016}. 298 K (green) and 464 K (pink) \cite{water_ixs_yamaguchi}). The corresponding \vus\s \cite{water_iapws} are represented by the triangles. (b) \sfast\s parameter, which is calculated from these two sound velocities, and (c) isochoric specific heat capacity \cvv\s \cite{water_iapws} are also presented. Curves are a guide for the eyes. (d) Contour plot of \sfast\s parameter as a function of temperature and pressure. Note the similarity to the contour map of \cvv\s (Fig.~\ref{fig:thermodynamics}(a))}
	\label{fig:water_fastsound_others}
\end{figure}

 In the low-temperature region, other researchers have reported estimates of sound velocity by IXS \cite{water_ixs_krisch2002} and INS (\vins) \cite{water_ins_ranieri2016} measurements at much higher pressures over 1 GPa.
 However, in these papers, they focused on the anomalous pressure variation of \vixs\s \cite{water_ixs_krisch2002} or microscopic dynamics \cite{water_ins_ranieri2016} and \sfast\s parameter was not deduced.
 We plot the values in Fig.~\ref{fig:water_fastsound_others} (a) by circles.
 The reported high-frequency sound velocities \vixs\s and \vins, which are indicated by circles, are significantly larger than the low-frequency sound velocity \vus\s \cite{water_iapws} indicated by the triangles, and have different pressure dependences.
 The \sfast\s parameter, which we calculated from the ratio of these two velocities, is shown in Fig.~\ref{fig:water_fastsound_others}(b);
 its pressure variations at approximately 300 K (277, 297, and 298 K) and 450 K (450 and 464 K) are different, but they are similar to those of \cvv\s \cite{water_iapws}, indicated in Fig.~\ref{fig:water_fastsound_others}(c).
 With increasing pressure, the \sfast\s parameter substantially decreases at approximately 300 K; in contrast, it also decreases but the variation is small at approximately 450 K.
 Figure~\ref{fig:water_fastsound_others}(d) plots the contour of the \sfast\s parameter on the temperature--pressure plane.
 It forms a large hill that seems to be topped in the low temperature (probably supercooled region) and low pressure region and extends over a wide range of temperature and pressure regions.
 Note that this figure is similar to the contour map of \cvv\s in the low temperature region (Fig.~\ref{fig:thermodynamics}(c)).
 It can be confirmed that the linkage between \sfast\s and \cvv\s is valid even in such a high-pressure region.
 In the end, the relaxation phenomenon measured in this study is well verified as the cause of the specific heat anomaly in liquid water.

\subsection{\label{ssec:thermodynamics} Pressure and temperature dependences of thermodynamic quantities}
 In liquid water, thermodynamic quantities in temperature and pressure ranges up to 1273 K and 1000 MPa, respectively, are expressed and can be obtained by an empirical formula, namely the IAPWS-95 formula, which was constructed based on a large amount of experimental data \cite{water_iapws}.
 The values from this equation are shown in Fig.~\ref{fig:water_params_full} and Fig.~\ref{fig:water_fastsound_others} presenting the experimental results.
 In this section, by observing contour maps (temperature and pressure dependences) of some quantities, we reiterate the anomalies in the thermodynamics of water and discuss the effect of LGT and LLT critical fluctuations.
 In particular, we will show that sound velocity, which is the focus of this study, is the key parameter in understanding such anomalies.
 Note that, in the original paper, only mathematical formula or tabulated values were listed and no contour maps were provided, which enables us to grasp the overview of thermodynamics.
 
 The first thermodynamic quantity to observe is the magnitude of density fluctuation, \szero.
\begin{figure*}
	\includegraphics[width=180mm]{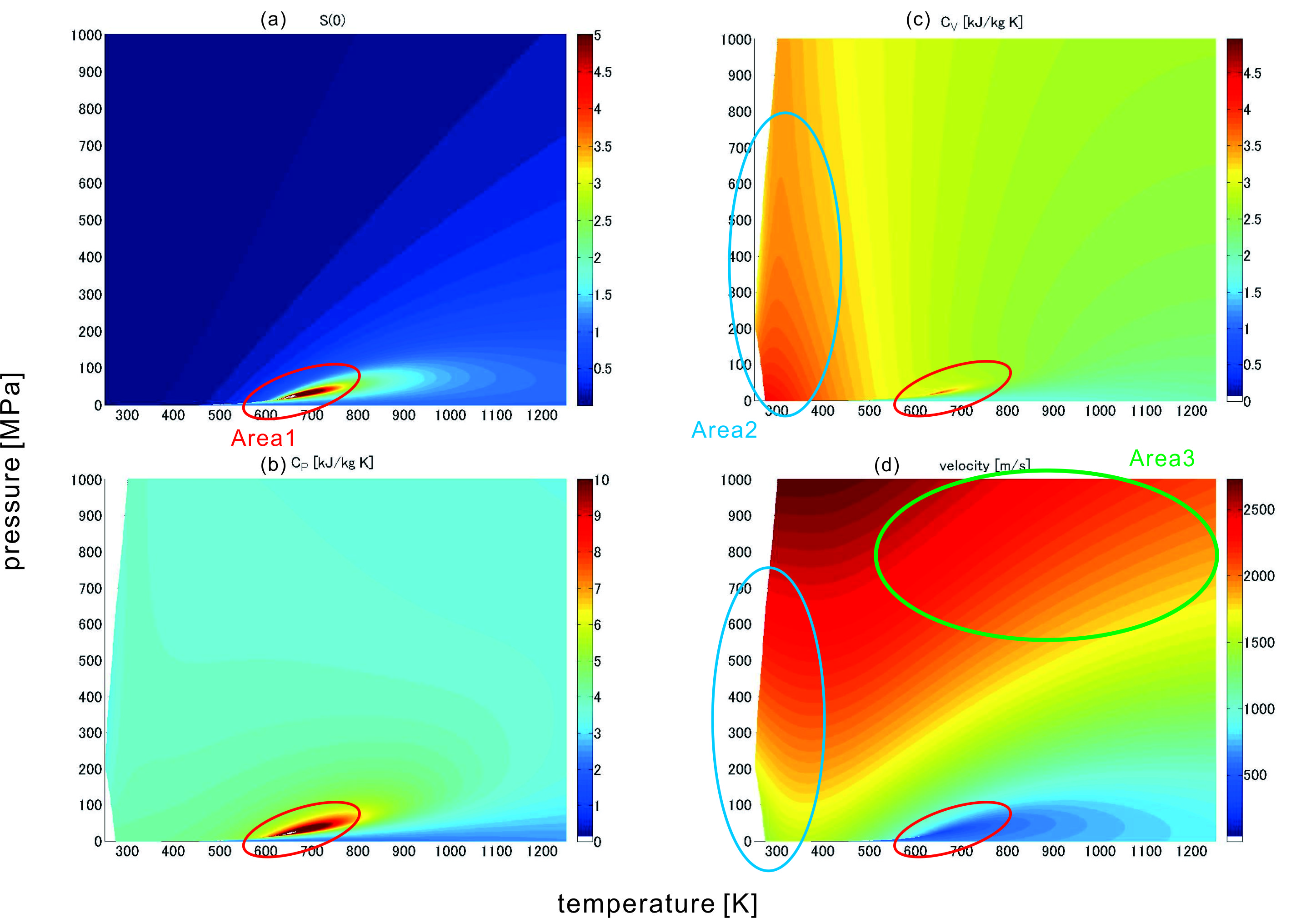}
	\caption{Contour maps of thermodynamic quantities calculated by the IAPWS-95 formula \cite{water_iapws}: (a) \szero, (b) \cpp, (c) \cvv, and (d) \vus. Areas 1, 2, and 3 indicated by the red, sky-blue, and green ellipses, represent the LGT critical, low-temperature-\cvv-hill, and ''normal'' sound velocity areas, respectively. }
	\label{fig:thermodynamics}
\end{figure*}
 Usually, in phase transitions of fluid systems such as LGT, critical fluctuation refers to density fluctuation.
 In Fig.~\ref{fig:thermodynamics}(a), there exists a thin and sharp mountain along the coexistence curve or critical ridge line in Area 1 indicated by a red eclipse.
 From this figure, the spread of the LGT critical fluctuation can be recognized, the top of which is the LGT critical point (647 K and 22 MPa).
 The next parameter that we will investigate is specific heat, the most important parameter in thermodynamics.
 There are two types of specific heat: isobaric specific heat capacity \cpp, and isochoric specific heat capacity \cvv, but the former is usually used because it is easier to measure experimentally.
 The contour map of \cpp\s is shown in Fig.~\ref{fig:thermodynamics}(b).
 As in \szero, a thin and sharp mountain spread out centered at the LGT critical point.
 \cpp\s can also be recognized as a good indicator of LGT critical fluctuation in conjunction with \szero.
 These two thermodynamic quantities are said to diverge toward the critical point with the same critical exponent $\gamma$ in the theory of critical phenomena \cite{book_stanley}.
 In contrast, observing these two contours, we notice that nothing anomalous can be recognized on this scale near ambient conditions.
 This, of course, does not indicate that there is no change whatsoever near ambient conditions;
 in fact, the well-known thermodynamic anomaly exists in ambient liquid water.
 The changes are overwhelmingly small.
 This is the main reason we concluded that these parameters are not appropriate for presenting the LLT critical fluctuation.
 This point will be further discussed in the next subsection.

 Meanwhile, the contour map of the isochoric specific heat capacity \cvv\s (Fig.~\ref{fig:thermodynamics}(c)) is significantly different from that of \cpp.
 Similar to \cpp, in Area 1, there exists a mountain topped by the LGT critical point, indicating that \cvv\s is a good indicator of LGT critical fluctuation.
 The height and spread of the mountain are not as great as in \cpp.
 Meanwhile, in the low-temperature region in Area 2, which is indicated by a blue eclipse, there is a widespread hill whose top seems to exist near ambient conditions (probably supercooled region), whose behavior is totally different from that of the mountain in the high-temperature region.
 Remarkably, this hill extends over a fairly high-temperature and high-pressure region up to 1000 MPa, which has a dominant effect on \cvv\s in nearly the left-half region of this figure.
 Apparently, the well-known anomaly of the specific heat capacity of liquid water under ambient conditions is due to this low-temperature-\cvv-hill.
 In other words, understanding the origin of this hill is the key to understanding the thermodynamics of liquid water, including the anomalies under ambient conditions.
 The contour map of (adiabatic or low-frequency) sound velocity (Fig.~\ref{fig:thermodynamics}(d)) is important in understanding this \cvv\s hill.
 Although there is no particular description about the feature of the ``speed of sound'' in the original paper \cite{water_iapws}, most of the original experimental data are obtained using the US method; 
 hence, it is reasonable to recognize this as US sound velocity \vus.
 Considering the LGT critical area (Area 1), it can be clearly observed that \vus\s decreases significantly around the LGT critical point compared to the surrounding area.
 This decrease is linked with the increase in \cvv\s seen in Fig.~\ref{fig:thermodynamics}(c). 
 Before looking at the low-temperature region, please look at Area 3 indicated by a green eclipse here, which is sufficiently high temperature and high-pressure regions.
 It is evident that the sound velocity changes almost linearly with both temperature and pressure, which is normal variation for liquids.
 Considering this, the behavior of the low temperature region Area 2 is clearly abnormal;
 the sound velocity is abnormally reduced.
 The spread of this reduction behavior is mostly consistent with the spread of \cvv\s as shown in Fig.~\ref{fig:thermodynamics}(c);
 i.e., the reduction of the sound velocity is linked to the increase in \cvv.
 The linkage between \cvv\s and reduction of (low-frequency) sound velocity is similar to the case of LGT critical fluctuation;
 We believe that it is a good reason to recognize it as LLT critical fluctuation, which will be discussed further in the next subsection.
 Here, we recognize the anomalous reductions of (low-frequency) sound velocity from the wide range of temperature--pressure dependence.
 In contrast, our (Fig.~\ref{fig:water_params_full}(a)) and others' high-frequency sound velocity obtained by IXS \cite{water_ixs_krisch2002} or INS \cite{water_ins_ranieri2016} (Fig.~\ref{fig:water_fastsound_others} (a)) show no such anomalous reductions and exhibits ``normal'' temperature--pressure variations, which is the key point revealed in this study;
 This implies that the reductions of sound velocity recognized in the contour map (Fig.\ref{fig:thermodynamics}(d)) are due to relaxation phenomena in the mesoscopic scales.
 In other words, the temperature--pressure dependence and time--space dependence of sound velocity of liquid water can be comprehensively understood by focusing on relaxation phenomena.
In the end, it was concluded that the thermodynamic anomalies of liquid water under ambient conditions are represented by \cvv-hill in the low temperature region, and its origin is a relaxation phenomenon of sound waves.

\subsection{\label{ssec:lgt_llt_difference} Essential difference between LGT and LLT critical fluctuations}
\begin{figure}[!h]
	\includegraphics[width=80mm]{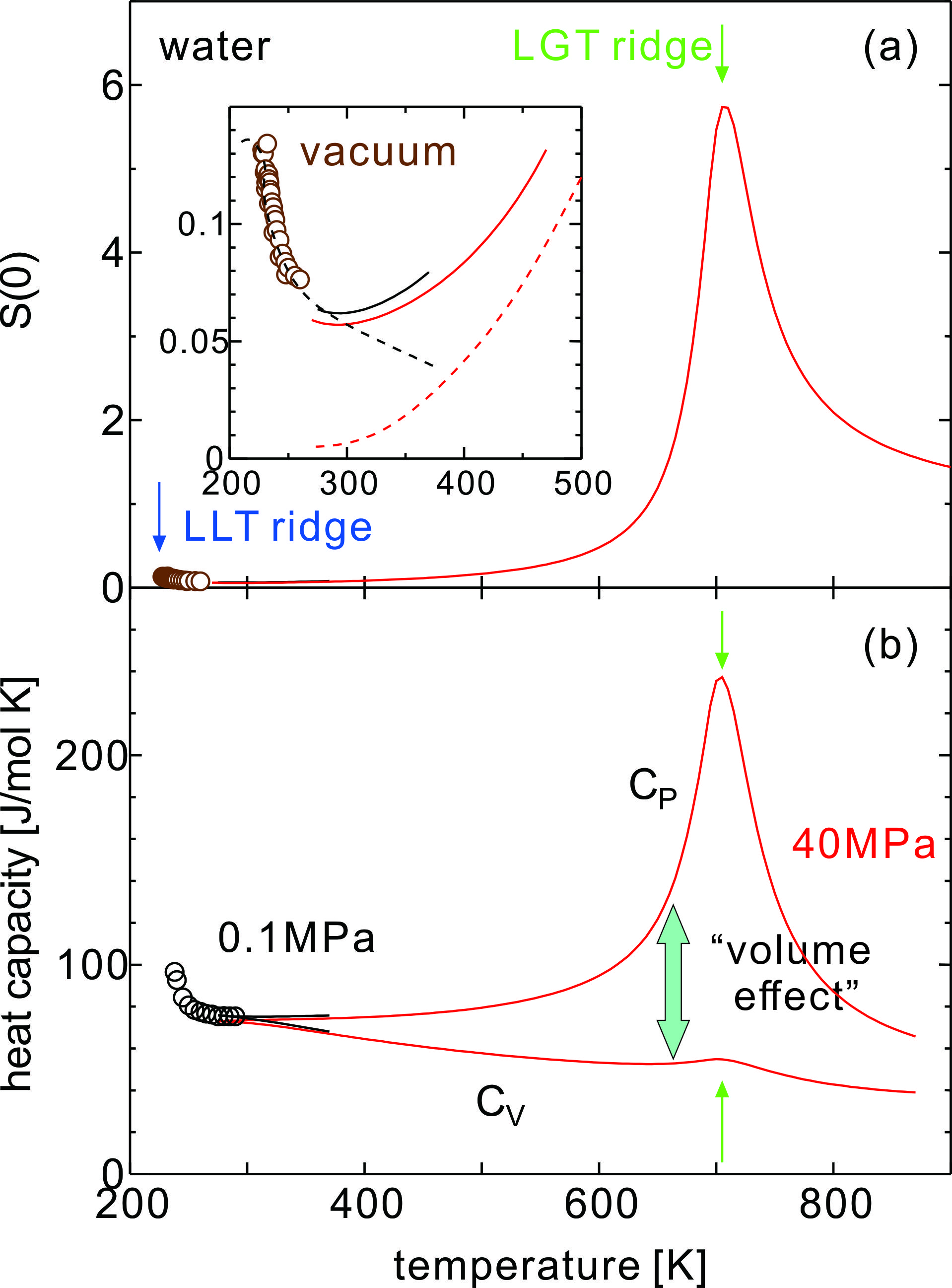}
	\caption{(a) \szero\s parameter of liquid water above melting temperature calculated using the IAPWS-95 formula \cite{water_iapws} are plotted by black (0.1 MPa) and red (40 MPa) lines. \szero\s values of supercooled liquid water obtained by SAS measurement at vacuum condition \cite{water_saxs_xfel} are also plotted by the brown circles.  Green and blue arrows indicate the peak positions of \szero\s value. (inset) Enlarged view in the low temperature region. The black and red dashed lines represent hypothetical \szero\s value due to LGT and LLT critical fluctuations. (b) Specific heat capacities \cpp\s and \cvv\s of liquid water calculated by IAPWS-95 formula \cite{water_iapws} are plotted by black (0.1 MPa) and red (40 MPa) lines. \cpp\s of supercooled liquid water obtained by experiment \cite{water_cp_supercool_angell} is also plotted by the circle. The difference between \cpp\s and \cvv\s is due to volume-effect. See eq.\ref{eq:dif_cpv}. }
	\label{fig:s0_cpv}
\end{figure}

 It seems reasonable to recognize the relaxation phenomena observed in the low temperature region as LLT critical fluctuation.
 However, there exists at least one major paradox;
 As shown in Fig.~\ref{fig:water_params_full}(b--d), while in LGT region, all the parameters of \sfast, \szero, \cpp, and \cvv\s were linked together, this linkage is not apparent in the low temperature region. 
 As mentioned in the main text, we believe that the essence of this problem is the large quantitative difference in density fluctuations between LGT and LLT.
 To explain this, the temperature dependence of  (a) \szero\s and (b) specific heat capacities (\cpp\s and \cvv) from near ambient conditions to LGT supercritical region are plotted in Fig.~\ref{fig:s0_cpv}.
 The pressures were 0.1 (black lines and circles), 40 MPa (red lines) and vacuum condition (brown circles).
 The lines are plotted using the IAPWS-95 formula \cite{water_iapws}, and the circles denote experimental values in the supercooled region \cite{water_saxs_xfel,water_cp_supercool_angell}.
 In particular, \szero\s is based on the experiment using XFEL \cite{water_saxs_xfel}, as discussed in the Introduction, and is claimed to have reached the critical ridge line of LLT, indicated by the blue arrow.
 In the figure, the difference in the pressure is not the essence of the discussion; 
 hence, we will only consider the temperature dependences.
 In the high-temperature region, \szero, \cpp, and \cvv\s all show their maximum around the LGT critical ridge line, indicated by green arrows at approximately 700 K.
 However, we can observe that the peak of \cvv\s is very small compared to that of the other two parameters.
 Meanwhile, we focus on the low-temperature region;
 The data in the supercooled region are naturally connected to the real liquid region above the melting temperature for both \szero\s and \cpp, and it is judged that there are no particular contradictions between the data in the supercooled region and those in the real liquid region.
 Moreover, the supercooled liquid is not considerably different from the real liquid.
 It is claimed that \szero\s exhibits its maximum value at the LLT critical ridge in the supercooled region (blue arrow); 
 however, it is observed that the absolute value is negligibly small compared with that at the LGT critical ridge line.
 This fact is an important point that has not received significant attention in previous LLT studies \cite{kajihara_saxs_review}.
 The ratio is $S(0)_\mathrm{LLT}/S(0)_\mathrm{LGT}=0.14/5.7=0.23\times10^{-2}$.
 The value can be understood with the following rough estimate.
 It can be assumed that in LGT, a liquid with a density of $\rho_{\mathrm{liq}}=1.0$ \gcm\s and a gas with a density of $\rho_{\mathrm{gas}}=0.0$ \gcm, and in LLT, a liquid with a low density of $\rho_{\mathrm{L}}=1.0$ \gcm\s and a liquid with a high density of $\rho_{\mathrm{H}}=1.1$ \gcm, become inhomogeneous; 
 thus, the order estimation of the ratio of these two \szero\s values for LGT and LLT can be expressed as follows:
\begin{eqnarray}
	\frac{S(0)_\mathrm{LLT}}{S(0)_\mathrm{LGT}} 
	\sim \frac{(\rho_{\mathrm{H}} - \rho_{\mathrm{L}})^{2}}{ (\rho_{\mathrm{liq}} - \rho_\mathrm{gas})^{2}} 
	\sim 10^{-2}.
	\label{eq:s0_mixmodel}
\end{eqnarray}
This value is the same order as the aforementioned \szero\s ratio, indicating that the \szero\s value at the supercooled LLT critical ridge is reasonable as the LLT critical density fluctuation. 
 Thus, such a comparison between LLT and LGT implies that there is an overwhelming quantitative difference in the density fluctuation even in the same critical fluctuations.
 This quantitative difference has a strong influence on the behavior of \szero\s near room temperature.
 The vicinity of room temperature is far from the LGT critical point; 
 however, there is an influence of the tail of the LGT critical density fluctuation that is 100 times that of the LLT critical density fluctuation.
 The influences of both these origins are considered, as shown in the inset of Fig.~\ref{fig:s0_cpv}(a) by the dashed lines. 
 It can be recognized that the influence of LLT density fluctuation (black dashed line) extends to a considerably high temperature region above melting temperature.
 This recognition is consistent with the fact that the population of the low-density liquid phase, which is estimated by X-ray emission spectroscopy, is large enough even at much higher temperatures than the melting point ($\simeq$0.30 at 350 K \cite{water_xes_xfel}).
 Near room temperature, \szero\s is small and no significant temperature change is observed. 
 However, this does not necessarily mean that there is no fluctuation or temperature variation.
 Both LLT and LGT critical fluctuations are substantially present, and the state in which the three phases of the low-density liquid, high-density liquid, and gas are mixed every moment with respect to the temperature changes.

Finally, we discuss the essential difference between LGT and LLT while focusing on \cpp\s and \cvv.
According to thermodynamics textbooks, \cpp\s and \cvv\s are enthalpy fluctuation (Appendix A in \cite{book_stanley}) and internal-energy fluctuation (p.130 in  \cite{book_statphys_kubo}), respectively, and are important parameters in discussing critical fluctuations.
As shown in Fig.~\ref{fig:s0_cpv}(b), \cpp\s and \cvv\s show large differences in absolute values and temperature dependences, and first focus on this difference:
The difference is expressed as follows (Eq. (2.20a) in \cite{book_stanley}),
\begin{eqnarray}
	C_{P} - C_{V} = T V \frac{\alpha_{P}^2}{\beta_{T}}.
	\label{eq:dif_cpv}
\end{eqnarray}
\alphap\s and \betat\s are thermal expansion coefficient and isothermal compressibility and represent the temperature and pressure derivative with respect to volume, respectively.
As mentioned above, the density difference between the two phases undergoing the phase transition is about 10 times larger in LGT than in LLT, and the above \alphap\s and \betat\s are also expected about 10 times larger in LGT than in LLT.
Therefore, the ``volume effect" \alphap$^2$/\betat\s that appears in the right-hand side of the equation is also expected to be about 10 times larger for LGT than for LLT.
Figure~\ref{fig:s0_cpv}(b) shows that the value \cpp$-$\cvv\s is very large in the LGT critical region around 700 K, indicating that the volume-effect due to such LGT critical fluctuation is very large.
In other words, such volume-effect account for most of the enthalpy fluctuation (\cpp) in LGT.
On the other hand, in the low temperature region, although there is no data on \cvv\s in the supercooled region, but \cpp$-$\cvv\s becomes almost zero near room temperature with cooling, indicating that the volume-effect in the LLT critical region is very small and almost negligible.
It is essential to consider this overwhelming difference in volume-effect when discussing heat capacities regarding LGT and LLT critical fluctuations.
Also with this in mind, it becomes possible to explain the difference in the temperature variations of \cpp\s and \cvv\s from 400 K to 600 K in Fig.~\ref{fig:s0_cpv}(b):
 The behavior of \cvv\s indicates that the internal energy fluctuation already increases from approximately 600 K with cooling; 
 However, in \cpp, the overwhelmingly large decrease in volume-effect can be regarded as masking the increasing characteristics of the internal energy fluctuation.
 What is important in LLT fluctuation may be seen as such internal energy fluctuation represented by \cvv.
 This implies that the essential difference between LGT and LLT is that LGT is a density (volume) phase transition while LLT is an internal energy phase transition.
The entity of the difference in internal energy in LLT was discussed in the previous Te paper \cite{kajihara_ixs_te}.
 In the case of Te, LLT is accompanied with MNMT, and the two phases of the transition are metal and nonmetal (semiconductor), and this difference in the valence electron's energy can be recognized as a large internal energy difference.
 In the case of Te, experimental results of its electrical properties indicate that the fluctuation or the two-fluidity of the metallic and nonmetallic states extends beyond the melting point to much higher temperatures and pressures, extending over almost the entire liquid region \cite{meltcurve_max_rapoport}.
 This fact is consistent with the existence of fast-sound \cite{kajihara_ixs_te} and large specific heat \cite{te_supercool} in the real liquid region above the melting temperature.
The existence of inhomogeneity in Te has been controversial for many years \cite{te_elect_cohen,te_inhomo_mott}, but its presence has been experimentally verified by SAS measurements, first in mixtures \cite{kajihara_saxs_sete,kajihara_saxs_review} and recently in pure Te \cite{te_supercool_saxs_peihao,kajihara_saxs_supercool_te_arxiv}.
Again, however, the mesoscopic fluctuations detected by density inhomogeneity in pure liquid Te become pronounced and exhibit maximum only in the supercooled region, the same as in the case of water \cite{water_saxs_xfel}.
 In the case of water, LLT is not accompanied by MNMT, of course.
However, experiments on X-ray emission \cite{water_xes_tokushima,water_xes_xfel,water_specta2016} and absorption spectroscopy \cite{water_xas_jpcm2002,water_xas_science2004_wernet,water_xas_science2004_smith} have proposed its two-fluid nature above melting temperature.
 It can be interpreted that even liquid water has two-fluid electronic states or internal energy fluctuation, and that the \cvv\s and \sfast\s parameters sensitively detect the corresponding internal energy fluctuation.
 With this perspective, the existence of fast-sound associated with MNMT of fluid Hg \cite{hg_fastsound} can be interpreted.
 The increase in density fluctuation due to MNMT is barely detectable \cite{kajihara_saxs_review}, but it is a three times fast-sound state and can be recognized as a large fluctuation state with a mixture of metallic and nonmetallic phases \cite{hg_mnmt_percolation_odagaki}.
 Interpreted in this way, the MNMT can be discussed comprehensively as a type of LLT.
 We concluded that there are essential differences between LGT and LLT:
 LGT is essentially a density phase transition while LLT is an internal energy (bonding state or valence-electronic state) phase transition;
 In conjunction with this, density fluctuation and internal energy fluctuation are important parameters for LGT and LLT critical fluctuations, respectively.
It is interesting to focus on \cvv\s and \sfast\s parameters for discussing LLT critical fluctuation.
We look forward to the development of new and unprecedented discussions regarding LLT in the future.
   

\bibliographystyle{apsrev4-2.bst}
\bibliography{D:/Literature/bibliography_kajihara}

\end{document}